# Differentially Private Search Log Sanitization with Optimal Output Utility

Yuan Hong[§], Jaideep Vaidya[§], Haibing Lu[†], and Mingrui Wu[‡]

[§]CIMIC, Rutgers University, NJ, USA, {yhong, jsvaidya}@cimic.rutgers.edu

[†]OMIS, Santa Clara University, CA, USA, hlu@scu.edu

[‡]Microsoft, Redmond, WA, USA, mingruiw@microsoft.com

## ABSTRACT

Web search logs contain extremely sensitive data, as evidenced by the recent AOL incident. However, storing and analyzing search logs can be very useful for many purposes (i.e. investigating human behavior). Thus, an important research question is how to privately sanitize search logs. Several search log anonymization techniques have been proposed with concrete privacy models. However, in all of these solutions, the output utility of the techniques is only evaluated rather than being maximized in any fashion. Indeed, for effective search log anonymization, it is desirable to derive the optimal (maximum utility) output while meeting the privacy standard. In this paper, we propose utility-maximizing sanitization based on the rigorous privacy standard of differential privacy, in the context of search logs. Specifically, we utilize optimization models to maximize the output utility of the sanitization for different applications, while ensuring that the production process satisfies differential privacy. An added benefit is that our novel randomization strategy ensures that the schema of the output is identical to that of the input. A comprehensive evaluation on real search logs validates the approach and demonstrates its robustness and scalability.

**Keywords:** Search Logs, Differential Privacy, Optimization

## 1. INTRODUCTION

Search engines are used by millions, if not billions, of people every day. The queries posed by the users form a large volume of data that can give great insight into human behavior via their search intent. Indeed, such data is invaluable for researchers and data analyzers in numerous fields [11]. For example, search engines themselves can use web search logs to identify common spelling errors, to recommend similar queries, or to expand queries. Many other applications also make use of search log data, such as the analysis of living habits from daily search, and the detection of epidemics [9]. For this reason, search log data is collected, stored, and analyzed in different ways by all search engines.

However, one problem with the storage and release of search log data is the potential for privacy breach. The queries that a user poses may sometimes reveal their most private interests and concerns. Thus, if search log data is published without sanitization or with trivial anonymization (such as simply replacing user ids by pseudonyms), many sensitive queries and clicks can be explicitly acquired by adversaries. [3, 11] demonstrates that it can only take a couple of hours to breach a particular user's privacy in the absence of good anonymization. Thus, it is crucial to anonymize search log data appropriately before storing or releasing it.

There has been significant work on database anonymization that looks at how to anonymize relational data. However, much of this work is not directly applicable since there are significant differences between search logs and relational data. Indeed, search logs pose additional challenges for anonymization. First, there is no explicit distinction between quasi-identifiers and sensitive information in search logs. Each user may pose hundreds of queries that involve lots of personal information (i.e. name, addresses, living habits, .etc) over a short period of time. By combining these queries, adversaries may easily discover an individual's identity, and it is difficult to foresee all possible combinations that can lead to privacy breaches. For instance, Table 1 illustrates a subset of an Internet user Alice's search log (note: the user-IDs can be determined by cookies, IP addresses or user accounts; we ignore query time and item rank of search logs in this paper). Although the real user-ID has been replaced by the pseudonymous ID 000101, the adversaries can still identify Alice's search log if they have some background knowledge on Alice (i.e. her address, she bought a second-hand Honda car via autotrader recently, she likes pizza), and thus learn more sensitive information (i.e. pregnancy test) from Alice's complete search log. Second, search logs are sparse and highly-dimensional, thus it is more difficult to guarantee rigorous privacy without sacrificing too much utility.

**Table 1: An Example of Search Logs**

| User-ID | Query | URL | Count |
|---|---|---|---|
| 000101 | 1 Washington Avenue | maps.google.com | 5 |
|  | Honda | www.honda.com | 2 |
|  | autotrader | www.autotrader.com | 4 |
|  | pizza | www.pizzahut.com | 1 |
|  | pregnancy test | www.medicinenet.com | 1 |
|  | ... | ... | ... |

In recent years, several search log anonymization techniques have been proposed in the literature to resolve the above problems [20, 5, 17, 14, 15, 19, 23]. Several anonymity models have been proposed for this domain along with corresponding anonymization algorithms. However, their basic premise is simply that the algorithm must satisfy the privacy requirements without worrying about the tradeoff between privacy and utility. Ideally, what is needed is a strategy that can maximize the utility while satisfying a given privacy requirement. To our knowledge, there is little work focusing on this challenging and practical problem. In this paper, we take the first step towards tackling this problem in the domain of search log anonymization by formulating utility-maximizing problems while ensuring a rigorous privacy standard.

## 1.1 Contribution

Given a particular privacy requirement, the utility-maximizing problem requires finding a way to anonymize search logs in a manner that satisfies the privacy standard and simultaneously achieves the optimal output utility. This requires deciding on a suitable privacy requirement as well as appropriate data utility measure. While several different anonymity models have been proposed in the literature, in this paper, we utilizes the robust privacy definition of differential privacy [7] (which lowers the privacy breach risk even if the adversaries hold arbitrary prior knowledge). We also define several different notions of utility and propose differentially private sanitization methods that can maximize the output utility. Thus, the main contributions of this paper are summarized as follows:

- The differentially private randomization in prior work (Korolova et al.[19] and Götz et al.[10]) ensures differential privacy by adding Laplacian noise to the aggregated query and clicked url counts. However, such approaches break the association between distinct query-url pairs in the output since all the user-IDs have been removed, which might be useful in only a few applications. Therefore, we propose differentially private algorithms based on a different randomization strategy: *sample user-IDs for every click-through query-url pairs using multinomial distribution*, which preserves user-IDs. This, to our knowledge, is the first randomization strategy to generate output with identical schema as the input search log. Thus, the sanitized search log can be analyzed in exactly the same fashion and for the same purpose as the input.

- Within our approach, the randomization algorithm also ensures the utility-maximized output that is still differentially private. To do this, we formally define the utility-maximizing problem: find an optimal sanitization that maximizes the output utility while satisfying differential privacy. Specifically, for quantifying the output utility, we define three different utility notions (measuring the utility of frequent click-through query-url pairs, the query-url pair diversity, etc.) that could benefit different applications (essentially, any utility measure can be coupled into our differentially private sanitization by replacing the utility objective function). We also prove that our sanitization satisfies differential privacy;

- We transform the utility-maximizing problems into standard optimization problems. We can now leverage prior developed effective solvers and adapt them to our problem. We experimentally validate the utility using real data sets.

The remainder of this paper is organized as follows. Section 2 reviews the related literature. In Section 3, we present our privacy model and the sanitization process. Section 4 introduces the constraints that guarantee privacy protection. We then formulate three different utility-maximizing problems and show that the corresponding sanitization methods are differentially private in Section 5. Section 6 evaluates the output utility of the proposed sanitization approaches. Finally, Section 7 concludes the paper.

## 2. RELATED WORK

### 2.1 Search Log Anonymization

Following the AOL search log incident, there has been some work on user privacy issues related to privately publishing search logs. Adar [1] proposes a secret sharing scheme where a query must appear at least $t$ times before it can be decoded. It may potentially remove too many harmless queries, thus reducing data utility. Kumar et al. [20] propose an approach that tokenizes each query tuple and hashes the corresponding search log identifiers. However, inversion cannot be done using just the token frequencies. Also, serious leaks are possible even when the order of tokens is hidden.

More recently, some anonymization models [19, 14, 15, 23] have been developed for search log release. He et al. [14], Hong et al. [15] and Liu et al. [23] anonymized search logs based on k-anonymity which is not as rigorous as differential privacy [10]. Korolova et al. [19] first applied the rigorous privacy notion – differential privacy to search log release by adding Laplacian noise. However, several shortcomings can be discovered in this work. First, the released result of this is the statistical information of queries and clicks where all users' search queries and clicks are aggregated together (without individual attribution). The data utility might be greatly reduced since the association between query-url pairs has been removed (the published data in Götz et al. [10] also suffers this constraint). With the released data, we cannot develop personalized query suggestion or recommendation for search engines, and also, we cannot carry out human behavior research since the output data do not include the information that any two queries belong to the same user. Second, as addressed by Götz et al. [10], the relaxed differential privacy notion in [19] is not sufficiently strong. Third, the utility in [19] is merely evaluated but not shown to be maximized. Adding Laplacian noise to the counts of selected queries and urls is straightforward and we cannot directly model optimization problems to maximize the output utility. Alternatively, our paper is to seek the maximum output utility for a novel differentially private search log sanitization mechanism which generate outputs with the identical schema as the original search log.

Furthermore, Götz et al. [10] analyzes algorithms of publishing frequent keywords, queries and clicks in search logs and conducts a comparison w.r.t. two relaxations of $\epsilon$-differential privacy (*relaxations are indispensable in search log publishing*). Our work utilizes the stronger relaxation of $\epsilon$-differential privacy – probabilistic differential privacy. Since we explore the optimal utility in our differentially private sanitization mechanism which outputs search logs rather than the results of counting queries and clicked urls over the search log, our work has a completely different focus, compared with their work [10].

### 2.2 Differential Privacy

In the context of relational data anonymization, Dwork et al.[6, 7] have proposed the rigorous privacy definition of differential privacy: a randomized algorithm is differentially private if for any pair of neighboring inputs, the probability of generating the same output, is within a small multiple of each other. This means that for any two datasets which are close to one another, a differentially private algorithm will behave approximately the same on both data sets. This notion provides sufficient privacy protection for users regardless of the prior knowledge possessed by the adversaries. This has been extended to data release in various different contexts besides search logs (i.e. contingency tables, graph data). Specifically, Xiao et al. [27] introduced a data publishing technique which ensures $\epsilon$-differential privacy while providing accurate answers for range-count queries. Hay et al. [13] presented an efficient algorithm for releasing a provably private estimate of the degree distribution of a network where it also satisfies the differential privacy. McSherry et al. [25] solved the problem of producing recommendations from collective user behavior while providing differential privacy for users. Our work follows the same line of research.

## 2.3 Tradeoff between Privacy and Utility

For any data modification based anonymization technique, a tradeoff between privacy and utility naturally holds. Li et al. [22] analyzed the fundamental characteristics of privacy and utility, and proposed a tradeoff framework for discussing privacy and utility. In microdata disclosure, Bayardo et al. [4] and LeFevre et al. [21] raised the optimal k-anonymity and the optimal multidimensional anonymization problem respectively. Kifer et al. [18] presented a way to gain additional utility from k-anonymous and l-diverse tables. Recently, Ghosh et al. [8] introduced a utility maximizing mechanism for releasing a statistical database. However, there is little work on this topic in the context of differential privacy guaranteed search log release. To our knowledge, we takes a first step towards addressing this deficiency.

## 3. MODEL

## 3.1 Differential Privacy

Our objective is to privately sanitize the input search logs that includes pseudonymous user-IDs, search queries, clicked urls and the counts of every user's click-through query-url pairs. Hence, we ensure that the output has the identical schema as the input: every single tuple in the output includes a pseudonymous user-ID, a click-through query-url pair and its count for this user.

We consider two search logs to be neighbors if they differ by an arbitrary user's (all) query tuples. Hence, we define every user's all query tuples in a search log $D$ as its user log.

DEFINITION 1. *(USER LOG $A_k$) Given a search log $D$, we denote each user $s_k$'s user log $A_k$ as all his/her query tuples in $D$, where every single tuple $[s_k, q_i, u_j, c_{ijk}] \in A_k$ includes a pseudonymous user-ID ($s_k$), a query ($q_i$), a url ($u_j$) and the count ($c_{ijk}$) of query-url pair $(q_i, u_j)$ belonging to user $s_k$.*

Clearly, every search log $D$ consists of numerous individual user logs ($D = \bigcup_{\forall s_k \in D} A_k$). Given two neighboring input search logs $D$ and $D'$ (w.o.l.g, $D = D' + A_k$), ensuring $\epsilon$-differential privacy for all the outputs might be impossible: for any output $O$ including items in $D$ but not in $D'$ (such as user-ID $s_k$), the probability that generating $O$ from $D'$ is zero but from $D$ is non-zero, hence the ratio between the probabilities cannot be bounded by $e^\epsilon$ (due to a zero denominator). We thus adopt the following relaxed notion of differential privacy (using our notations):

DEFINITION 2. *(($\epsilon, \delta$)-PROBABILISTIC DIFFERENTIAL PRIVACY [24, 10]) A randomization algorithm $\mathcal{R}$ satisfies ($\epsilon, \delta$)-probabilistic differential privacy if for any input search log $D$, we can divide the output space $\Omega$ into two sets $\Omega_1, \Omega_2$ such that*
*(1) $Pr[\mathcal{R}(D) \in \Omega_1] \leq \delta$, and*
*for $D$'s all neighboring search logs $D'$ and for any output $O \in \Omega_2$:*
*(2) $\frac{Pr[\mathcal{R}(D)=O]}{Pr[\mathcal{R}(D')=O]} \leq e^\epsilon$ and $\frac{Pr[\mathcal{R}(D')=O]}{Pr[\mathcal{R}(D)=O]} \leq e^\epsilon$.*

The above probabilistic differential privacy ensures that $\mathcal{R}$ satisfies $\epsilon$-differential privacy with high probability (no less than $1 - \delta$) [10]. In this definition, the set $\Omega_1$ includes all privacy-breaching outputs for $\epsilon$-differential privacy where the probability of generating such outputs is bounded by $\delta$. Specifically in our sanitization (w.o.l.g. $D = D' + A_k$), since we retain *user IDs* in the output and $D'$ does not contain $s_k$, we can only consider $\Omega_1$ as the output space where all outputs in $\Omega_1$ include user-ID $s_k$ (because $\epsilon$-differential privacy cannot be achieved when $D', D$ differing in user $s_k$'s user log $A_k$ and the output $O$ including $s_k$). Hence, the probability $Pr[\mathcal{R}(D) \in \Omega_1]$ should be no greater than $\delta$ (the probability of $s_k$ existing in the overall output space $\Omega$ should be bounded by $\delta$). Moreover, for any output $O \in \Omega_2$, two ratios should be bounded by $e^\epsilon$ for achieving $\epsilon$-differential privacy. Definition 2 has been proven to be stronger than the privacy notion of Korolova et al.'s work [19] (indistinguishability differential privacy [6]) by Götz et al.[10] (as also shown in Section 4.3).

All the sanitization methods addressed in this paper are required to satisfy this robust and rigorous privacy definition. No matter how much prior knowledge is owned by adversaries, we can lower the privacy risk by bounding the probabilities that any arbitrary two neighboring inputs produce any possible output.

## 3.2 Search Log Sanitization Process

With a rigorous privacy standard (Definition 2), our goal is to maximize the retained utility for the sanitized search logs. We now illustrate our search log sanitization process that integrates the satisfaction of differential privacy and utility maximization.

The most sensitive values in search logs are the click-through information. Sometimes search queries may be more sensitive than the clicked urls in search logs (i.e. query "diabetes medicine" and click "www.walmart.com"), or vice versa (i.e. query "medicine" and click "www.cancer.gov"). We thus consider each distinct click-through query-url pair (simply denoted as query-url pair) as a combination of the sensitive values in the search logs. In our privacy model, Definition 2 ensures that adding any user's all search information (user-ID, query-url pairs and the counts) in the input does not cause any additional risk.

Table 2 presents some frequently used notations in our model: we denote $c_{ij}$ as the input count of any query-url pair $(q_i, u_j)$ and the set of these counts $\{\forall c_{ij}\}$ constitutes the *input query-url histogram*. Similarly, $x_{ij}$ represents the output count of $(q_i, u_j)$ and the set of these counts $x = \{\forall x_{ij}\}$ forms the *output query-url histogram*. Finally, the output counts of all triplets $(q_i, u_j, s_k)$ form the *output query-url-user histogram* which is randomly sampled (the sampling process will be given later on). Similarly, the deterministic counts of all triplets $(q_i, u_j, s_k)$ in the input form the *input query-url-user histogram*.

**Table 2: Frequently Used Notations**

| | |
|---|---|
| $(q_i, u_j)$ | an arbitrary query-url pair in the input/output |
| $(q_i, u_j, s_k)$ | any user $s_k$'s arbitrary query-url pair $(q_i, u_j)$ |
| $c_{ij}$ | the total count of $(q_i, u_j)$ in the input |
| $c_{ijk}$ | the count of triplet $(q_i, u_j, s_k)$ in the input |
| $x_{ij}$ (variable) | the total count of $(q_i, u_j)$ in the output (in the optimal solution: $x_{ij}^*$) |
| $x_{ijk}$ (random variable) | the count of triplet $(q_i, u_j, s_k)$ in a sample output ($x_{ijk}^*$ is the count of $(q_i, u_j, s_k)$ if sampling with $x_{ij}^*$ trials) |

Algorithm 1 illustrates two steps of our sanitization. We first compute the optimal output counts for all the query-url pairs in the input search log $D$, and then generate the output $O$ by sampling user-IDs for each of them with multinomial distribution [2] (the details of this multinomial sampling are given later on). More specifically, the algorithm can be guaranteed to be differentially private by some constraints for the output counts of all query-url pairs $\{\forall x_{ij}\}$ (we can derive the constraints from the randomization, as shown in Section 4). Meanwhile, the output utility can be maximized by the utility objective function (some options are given in Section 5). Thus, we can formulate the utility-maximizing problem to compute the optimal output counts of all query-url pairs for the random sampling (the optimal solution $x^* = \{\forall x_{ij}^*\}$ achieves the optimal output utility and also satisfies differential privacy constraints).

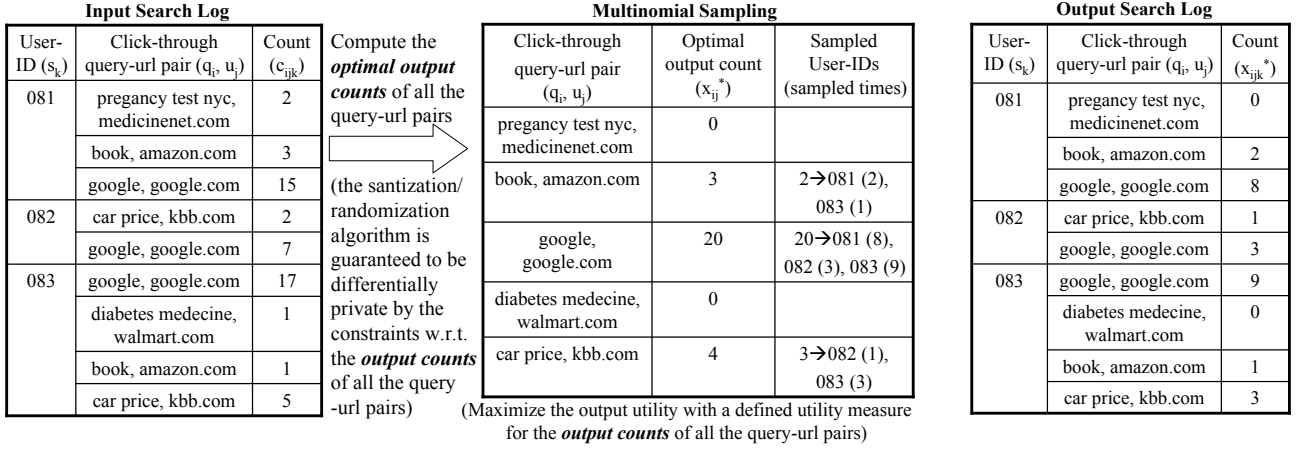

**Figure 1: An Example of the Sanitization Algorithm**

---

**Algorithm 1** Sanitization Algorithm

**Input:** search log $D$ and differential privacy parameters $(\epsilon, \delta)$
**Output:** sanitized search log $O$
1: **Compute the Optimal Output Counts** for all query-url pairs in the search log: $\{\forall (q_i, u_j) \in D, x_{ij}^*\}$.
   /*** solve an optimization problem: define a utility objective function w.r.t. the output counts $\{\forall x_{ij}\}$ while $\{\forall x_{ij}\}$ subject to some constraints that ensures differential privacy for this algorithm. (the optimal solution is $\{\forall x_{ij}^*\}$) ***/
2: **Generate the Output** $O$: sampling user-IDs for every query-url pair $(q_i, u_j)$ with $x_{ij}^*$ times multinomial trials (the probability of every sampled outcome in one trial is given by the input $D$).

---

Figure 1 shows an example of Algorithm 1, particularly the multinomial sampling after computing the optimal output counts of all query-url pairs $\{\forall x_{ij}^*\}$ (assume that $\{0, 3, 20, 0, 4\}$ in the example is the optimal solution of an optimization problem that includes a utility objective and some constraints ensuring differential privacy). Therefore, our multinomial sampling has following properties:

1. The number of multinomial trials for $(q_i, u_j)$'s user-ID sampling is given as $x_{ij}^*$ (optimal solution $x^* = \{\forall x_{ij}^*\}$).

2. In every multinomial trial for any query-url pair $(q_i, u_j)$, the probability that any user-ID $s_k$ is sampled, is $c_{ijk}/c_{ij}$. Specifically, i.e. "car price, kbb.com" in Figure 1, the probability that user 082 is sampled is $\frac{2}{0+2+5}$. However, the probability that user 081 is sampled for this query-url pair is 0. In addition, the expected value of every random variable $x_{ijk}$ can be derived as $E(x_{ijk}) = x_{ij} \cdot \frac{c_{ijk}}{c_{ij}}$. Thus, given an output count $x_{ij}^*$ (optimal) for any query-url pair $(q_i, u_j)$, the shape of the input/output query-url-user histograms w.r.t. only query-url pair $(q_i, u_j)$ (illustrating the individual counts of $(q_i, u_j)$ held by distinct users) should be analogous (this is guaranteed by multinomial distribution). i.e. the input/output query-url-user histogram w.r.t. "google, google.com", even if the output count $x_{ij}^* = 20 < c_{ij} = 15 + 7 + 17 = 39$, the shape of histograms $\{8, 3, 9\}$ (in a randomized output, see Figure 1(b)) and $\{15, 7, 17\}$ (in the input) is similar.

3. If $\forall (q_i, u_j)$, the *Input Support* (denoted as $c_{ij}/\sum_{\forall (q_i, u_j)} c_{ij}$), is close to the *Output Support* (denoted as $x_{ij}/\sum_{\forall (q_i, u_j)} x_{ij}$),

*the shape of the output query-url histogram* can be maximally preserved. At this time, after sampling user-IDs with the above output counts of all query-url pairs (or called output query-url histogram), *the shape of the output query-url-user histogram* can be maximally preserved as well.

Actually, one of our utility-maximizing problems is to seek the optimal output utility that minimizes the sum of the support distances for all frequent query-url pairs (see the definition and details in Section 5.2, if pursuing the minimum sum of support distances for all query-url pairs, we can lower the *minimum support threshold*). Thus, once the sum of the support distances are minimized (utility-maximizing problem can do so, i.e. it figures out that the distance between $\{\forall \frac{x_{ij}}{\sum x_{ij}}\} = \{0, \frac{3}{27}, \frac{20}{27}, 0, \frac{4}{27}\}$ and $\{\forall \frac{c_{ij}}{\sum c_{ij}}\} = \{\frac{2+0+0}{53}, \frac{3+0+1}{53}, \frac{15+7+17}{53}, \frac{0+0+1}{53}, \frac{0+2+5}{53}\}$ is minimized while satisfying some privacy guarantee constraints), the shape of the input/output query-url-user histograms can be analogous (i.e. see the counts in the left table of Figure 1(a) and Figure 1(b)).

To sum up, if we compute the output count of every query-url pair $x = \{\forall x_{ij}\}$ by solving an optimization problem (for variables $x = \{\forall x_{ij}\}$) that maximizes the output utility and also ensures differential privacy for the sanitization algorithm, the output with optimal utility can be generated by sampling user-IDs for all the query-url pairs (*the schema of Input/Output is indeed identical since we can sort the output by the sampled user-IDs, as shown in Figure 1(b) where the association between query-url pairs and the shape of query-url-user histogram can be preserved*).

## 4. PRIVACY GUARANTEE CONDITIONS

Assume that $\mathcal{R}$ is a sanitization algorithm that samples user-IDs for every query-url pair $(q_i, u_j)$ with its total output count $x_{ij}$. Since the sampling procedures for all query-url pairs are independent, for any input $D$ ($\{\forall c_{ijk}\}$ is given) and a possible output $O$ ($\{\forall x_{ijk}\}$ is also given), the probability $Pr[\mathcal{R}(D) = O]$ can be computed in terms of the probability mass function of multinomial distribution [2]:

$$Pr[\mathcal{R}(D) = O] = \prod_{\forall (q_i, u_j) \in O} [x_{ij}! \cdot \prod_{\forall s_k \in D} \frac{(c_{ijk}/c_{ij})^{x_{ijk}}}{x_{ijk}!}] \quad (1)$$

Indeed, $Pr[\mathcal{R}(D) = O]$ is determined by $x_{ij}$ and $\{\forall s_k \in D, \frac{c_{ijk}}{c_{ij}}$ and $x_{ijk}\}$. Given input $D$, $\{\forall s_k, \frac{c_{ijk}}{c_{ij}}\}$ are constants.

Hence, if $\forall (q_i, u_j) \in D$, the output count $x_{ij}$ is determined, we can compute the probability $Pr[\mathcal{R}(D) = O]$ for any output $O \in \Omega$ ($\forall x_{ijk}$ are fixed in $O$). Therefore, given any pair of neighboring inputs $D$ and $D'$ that differ in one user log, bounding the probabilities per Definition 2 for a divided output space $\Omega$ can be transformed to the problem: determining a feasible solution $x = \{\forall x_{ij}\}$ in the output that satisfies all the probability bounding conditions in Definition 2 for an output space split $\Omega = \Omega_1 \cup \Omega_2$. Using this we can formulate the constraints (satisfying differential privacy) for variables: the counts of all query-url pairs $x = \{\forall x_{ij}\}$ in all the possible outputs $O \in \Omega$.

Without loss of generality, we let $D = D' + A_k$ where $D$ and $D'$ differ in an arbitrary user $s_k$'s user log $A_k$. Thus, we first derive the probabilities in Definition 2 for all $O$ in the output space $\Omega$, and then deduce the constraints for satisfying differential privacy.

## 4.1 Probabilities in Definition 2

Due to $D = D' + A_k$, the user-ID $s_k$ might be sampled into the output $O$ if starting from $D$. Thus, for all outputs $O$ which contain $s_k$, we have $Pr[\mathcal{R}(D') = O] = 0$ (since $s_k \notin D'$). Recall that, given $A_k = D - D'$ (or $A_k = D' - D$), we can only divide the output space $\Omega$ into two sets $\Omega_1$ and $\Omega_2$ as: (1) every output $O$ in $\Omega_1$ includes $s_k$; (2) every output $O$ in $\Omega_2$ does not include $s_k$, because $\Omega_1$ should includes all the exceptional outputs that violates $\epsilon$-differential privacy. We thus bound the probabilities per Definition 2 for the above output space split of any two neighboring inputs ($\forall O \in \Omega_1$, user-ID $s_k \in O$ and $s_k \notin \Omega_2$) to achieve differential privacy.

### 4.1.1 for all $O \in \Omega_1$

Since $\forall O \in \Omega_1$ where $s_k \in O$, we have $Pr[\mathcal{R}(D') = O] = 0$. Thus, the probability $Pr[\mathcal{R}(D') \in \Omega_1]$ is also equal to 0. We now compute the probability $Pr[\mathcal{R}(D) \in \Omega_1]$.

Specifically, to generate any possible output $O$ including user-ID $s_k$ from $D$, the probability $Pr[\mathcal{R}(D) = O]$ (where $O \in \Omega_1$) is equal to the probability that "$s_k$ is sampled at least once in the multinomial sampling process of all the query-url pairs in $A_k$". For every query-url pair $(q_i, u_j) \in A_k$, if its total output count in the sampling is $x_{ij}$, the probability that $s_k$ is not sampled in a single multinomial trial (a user-ID in $D$ except $s_k$ is sampled) is $\frac{c_{ij}-c_{ijk}}{c_{ij}}$ simply because user $s_k$ holds $(q_i, u_j)$ with the count $c_{ijk}$ and the total count of $(q_i, u_j)$ is $c_{ij}$ in the input $D$. Since $\forall (q_i, u_j) \in A_k$ may lead to that $s_k$ being sampled and the multinomial sampling for every query-url pair $(q_i, u_j)$ includes $x_{ij}$ independent trials, we have $Pr[s_k \text{ is not sampled}] = \prod_{\forall (q_i, u_j) \in A_k} (\frac{c_{ij}-c_{ijk}}{c_{ij}})^{x_{ij}}$. Finally, we can obtain the probability that $s_k$ is sampled at least once: $Pr[s_k \text{ is sampled}] = 1 - \prod_{\forall (q_i, u_j) \in A_k} (\frac{c_{ij}-c_{ijk}}{c_{ij}})^{x_{ij}}$.

Thus, we can derive the probability $Pr[\mathcal{R}(D) \in \Omega_1]$ as below:

$$Pr[\mathcal{R}(D) \in \Omega_1] = 1 - \prod_{\forall (q_i, u_j) \in A_k} (\frac{c_{ij}-c_{ijk}}{c_{ij}})^{x_{ij}} \quad (2)$$

One important issue is worth noting in multinomial sampling. For any query-url pair $(q_i, u_j) \in A_k$ where $c_{ijk} = c_{ij}$ ($(q_i, u_j)$ is unique and only belongs to user $s_k$), if its output count $x_{ij} > 0$, the probability $Pr[\mathcal{R}(D) \in \Omega_1]$ should be equal to 1 which cannot be bounded. Therefore, we let $x_{ij} = 0$ for this case and all the unique query-url pairs in the input should be removed.

### 4.1.2 for all $O \in \Omega_2$

For any output $O \in \Omega_2$, we discuss the ratios $\frac{Pr[\mathcal{R}(D)=O]}{Pr[\mathcal{R}(D')=O]}$ and $\frac{Pr[\mathcal{R}(D')=O]}{Pr[\mathcal{R}(D)=O]}$ (since $O$ does not include $s_k$, we have $Pr[\mathcal{R}(D) = O] > 0$ and $Pr[\mathcal{R}(D') = O] > 0$).

Intuitively, for all query-url pairs that belong to both $A_k$ and $D'$, sampling user-IDs from $D$ involves an additional candidate $s_k$ (but $s_k \notin O$) compared with sampling user-IDs from $D'$. We thus have $\frac{Pr[\mathcal{R}(D)=O]}{Pr[\mathcal{R}(D')=O]} \leq 1$ and $\frac{Pr[\mathcal{R}(D')=O]}{Pr[\mathcal{R}(D)=O]} \geq 1$. Since the ratio $\frac{Pr[\mathcal{R}(D)=O]}{Pr[\mathcal{R}(D')=O]}$ is bounded by 1 (and obviously $e^\epsilon$), we only need to derive the ratio $\frac{Pr[\mathcal{R}(D')=O]}{Pr[\mathcal{R}(D)=O]}$.

As mentioned in Section 4.1.1, all the query-url pairs in $D$ (and $A_k$) but not in $D'$ should be not be retained in the output. Thus, to generate $O$ from $D$, we only sample user-IDs for the common query-url pairs of $D$ and $D'$. Two categories of common query-url pairs can be identified in $D'$ ($D' \subset D$ here): (1) query-url pairs in $D'$ but not in $A_k$ (2) query-url pairs in $D'$ and also in $A_k$.

In the first category, $\forall (q_i, u_j)$ in $D'$ but not in $A_k$, the probabilities of sampling user-IDs for $(q_i, u_j)$ from $D$ and $D'$ are equivalent because the query-url-user histogram w.r.t. these query-url pairs in $D$ and $D'$ is identical. We denote the ratio of these two probabilities as $\frac{Pr[\mathcal{R}(D')=O]}{Pr[\mathcal{R}(D)=O]}(ij)$ that is equal to 1.

In the second category, $\forall (q_i, u_j)$ in $D'$ and also in $A_k$, we can consider every sampled user-ID in the process of $\mathcal{R}(D) \to O$ into two cases: "$s_k$ is sampled or not". In every multinomial trial for $(q_i, u_j)$, the probability of sampling $s_k$ is $\frac{c_{ijk}}{c_{ij}}$ while the probability of sampling another user-ID in $D$ (also in $D'$) is $1 - \frac{c_{ijk}}{c_{ij}}$. Since the number of $(q_i, u_j)$ in the output is $x_{ij}$ ($x_{ij}$ times independent trials), we have ratio $\frac{Pr[\mathcal{R}(D')=O]}{Pr[\mathcal{R}(D)=O]}(ij) = \frac{1}{(1-\frac{c_{ijk}}{c_{ij}})^{x_{ij}}} = (\frac{c_{ij}}{c_{ij}-c_{ijk}})^{x_{ij}}$ (since $O$ does not contain $s_k$, $s_k$ should not be sampled in $x_{ij}$ times independent trials when generating $O$ from $D$).

In sum, to generate any output $O \in \Omega_2$ from $D$ and $D'$ respectively, it is independent to sample user-IDs for all the above two categories of query-url pairs. Thus, $\forall O \in \Omega_2$, $\frac{Pr[\mathcal{R}(D')=O]}{Pr[\mathcal{R}(D)=O]} = \prod_{\forall (q_i, u_j) \in D'} \frac{Pr[\mathcal{R}(D')=O]}{Pr[\mathcal{R}(D)=O]}(ij)$. Since $\forall (q_i, u_j) \in D'$ but $\notin A_k$, $\frac{Pr[\mathcal{R}(D')=D]}{Pr[\mathcal{R}(D)=D]}(ij) = 1$, we have $\forall O \in \Omega_2$:

$$\frac{Pr[\mathcal{R}(D') = O]}{Pr[\mathcal{R}(D) = O]} = \prod_{\forall (q_i, u_j) \in D' \cap A_k} (\frac{c_{ij}}{c_{ij}-c_{ijk}})^{x_{ij}} \quad (3)$$

## 4.2 Differential Privacy Constraints

$(\epsilon, \delta)$-probabilistic differential privacy (Definition 2) demands: for any input $D$, $Pr[\mathcal{R}(D) \in \Omega_1] \leq \delta$; for $D$'s arbitrary neighboring input $D'$ and $\forall O \in \Omega_2$, $1/e^\epsilon \leq \frac{Pr[\mathcal{R}(D)=O]}{Pr[\mathcal{R}(D')=O]} \leq e^\epsilon$. We now show that proving the randomization algorithm to be $(\epsilon, \delta)$-probabilistic differentially private as per Definition 2 is equivalent to ensuring that the output counts of all query-url pairs satisfy a set of conditions. Theorem 1 is proven in Appendix A.

THEOREM 1. *The randomization algorithm $\mathcal{R}$ achieves $(\epsilon, \delta)$-probabilistic differential privacy if for any input search log $D$, the output counts of query-url pairs $x = \{\forall (q_i, u_j) \in D, x_{ij}\}$ satisfy:*

1. *if $\exists$ triplet $(q_i, u_j, s_k) \in D$ such that $c_{ijk} = c_{ij}$, then $x_{ij} = 0$ (do not output unique query-url pairs);*

2. *for all $A_k \subset D$: $\prod_{\forall (q_i, u_j) \in A_k} (\frac{c_{ij}}{c_{ij}-c_{ijk}})^{x_{ij}} \leq e^\epsilon$;*

3. *for all $A_k \subset D$: $1 - \prod_{\forall (q_i, u_j) \in A_k} (\frac{c_{ij}-c_{ijk}}{c_{ij}})^{x_{ij}} \leq \delta$.*

As a result, we can utilize these conditions to formulate utility-maximizing problems in our differentially private search log sanitization. Specifically, we can implement Condition 1 while preprocessing the input search log (removing all the unique query-url pairs), and regard Condition 2 and 3 as *Differential Privacy Constraints* in the sanitization. As soon as they are satisfied, the sanitization should be $(\epsilon, \delta)$-probabilistic differential private for every pair of neighboring search logs that differs in only one user log.

Note that while our multinomial sampling process is differentially private, the computation of the counts ($x^* = \{\forall x_{ij}^*\}$) is not necessarily so. To make the whole (end-to-end) sanitization differentially private, we must ensure that the count computation step is also differentially private. One simple way to do this is to use the generic procedure of adding Laplacian noise to the counts derived from the optimization model ($x^* = \{\forall x_{ij}^*\}$). Since the count computation can be viewed as a query over the input database, adding Laplacian noise will make the computation differentially private.

Specifically, similar to Korolova et al. [19], if the count differences of every query-url pair $(q_i, u_j)$ in the optimal solutions derived from two neighboring inputs $(D, D')$ are bounded by a constant $d$, computing optimal counts can be guaranteed to be $\epsilon'$-differentially private [19] ($\epsilon'$ is the parameter of ensuring differential privacy for such step) by adding Laplacian noise to the optimal count of every query-url pair: $\forall (q_i, u_j), x_{ij}^* \leftarrow x_{ij}^* + Lap(d/\epsilon')$. Essentially, given $d$, we can simply bound the difference of every query-url pair's optimal count (computed from any two neighboring inputs $D, D'$) by executing the following preprocessing procedure for every user log $A_k$ in the input database ($D$ or $D'$):

1. formulate two utility-maximizing problems (pick the same option as the following sanitization) with neighboring inputs $D$ and $D - A_k$ (or $D'$ and $D' - A_k$ if $D'$ is the input) respectively, and solve them.

2. if the count difference of any query-url pair in both optimal solutions is greater than $d$, remove $A_k$ from $D$ (or $D'$) [1].

If applying the above preprocessing procedure to any two neighboring inputs $D$ and $D'$, and computing the optimal output counts with the updated $D$ and $D'$, the difference of every query-url pair's optimal count can be bounded by $d$. Thus, adding noise $Lap(d/\epsilon')$ can ensure $\epsilon'$-differential privacy [19] for the step of computing optimal counts in Algorithm 1. While adding noise may distort the optimality to some extent, this is the price of guaranteeing complete differential privacy. Since adding Laplacian noise is a well-studied generic approach, we do not discuss this differential privacy guarantee due to space limitation, and <u>the sanitization/randomization algorithm refers to the sampling process in this paper</u>.

### 4.3 Indistinguishability Differential Privacy

Recall that in Section 3.1, we have noted that probabilistic differential privacy [24, 10] provides stronger privacy guarantee than indistinguishability differential privacy [6, 19]. Particularly, the probabilistic differential privacy notion has following property:

PROPOSITION 1. *Probabilistic differential privacy implies indistinguishability differential privacy in our search log sanitization: if all the conditions in Definition 2 are satisfied with parameters $(\epsilon, \delta)$, the following two inequalities also hold:*

1. $Pr[\mathcal{R}(D') \in \widehat{O}] \leq e^\epsilon \cdot Pr[\mathcal{R}(D) \in \widehat{O}] + \delta$;

2. $Pr[\mathcal{R}(D) \in \widehat{O}] \leq e^\epsilon \cdot Pr[\mathcal{R}(D') \in \widehat{O}] + \delta$.

*where $\widehat{O}$ is an arbitrary set of possible outputs and $\widehat{O} \subseteq \Omega$.*

Götz et al. prove Proposition 1 and show that the converse of it does not hold in [10] (The proof of Proposition 1 is also given in Appendix B). Hence, satisfying Definition 2 with the differential privacy constraints (Theorem 1) provides more rigorous privacy guarantee than the work of Korolova et al. [19].

## 5. UTILITY-MAXIMIZING PROBLEMS

While search logs consist of millions of queries and click-through urls, from the perspective of utility, clearly, all are not equal. Indeed, from an application perspective, only a small portion may be useful with regards to a specific purpose. For instance, only the frequent query-url pairs are useful for query recommendation. Hence, different data usage purposes may result in different requirements for extracting data from the original search log. To privately sanitize search logs while retaining maximal utility, we need to evaluate the data utility according to the usage requirement. In this section, we introduce three utility-maximizing problems with three different utility definitions.

### 5.1 Maximizing the Output Size

Before formulating the utility-maximizing problems, we first present the differential privacy constraints. As stated in Theorem 1, our sanitization algorithm satisfies $(\epsilon, \delta)$-probabilistic differential privacy if three conditions for the output counts of all query-url pairs are satisfied. Specifically, Condition 1 should be implemented in the preprocessing step[2] while Conditions 2 and 3 give two sets of constraints for the output counts of all query-url pairs, $x = \{x_{ij}\}$:

$$s.t. \begin{cases} \forall A_k \subset D, \prod_{\forall (q_i, u_j) \in A_k} (\frac{c_{ij}}{c_{ij} - c_{ijk}})^{x_{ij}} \leq e^\epsilon \\ \forall A_k \subset D, 1 - \prod_{\forall (q_i, u_j) \in A_k} (\frac{c_{ij} - c_{ijk}}{c_{ij}})^{x_{ij}} \leq \delta \\ \forall x_{ij} \geq 0 \text{ and } x_{ij} \text{ is an integer} \end{cases}$$

Intuitively, the differential privacy constraints can be transformed into **linear constraints**: (constant $t_{ijk} = \frac{c_{ij}}{c_{ij} - c_{ijk}}$; each user log $A_k$'s two constraints can be combined as $\min\{\epsilon, \log \frac{1}{1-\delta}\}$)

$$s.t. \begin{cases} \forall A_k \subset D, \sum_{\forall (q_i, u_j) \in A_k} x_{ij} \cdot \log t_{ijk} \leq \min\{\epsilon, \log \frac{1}{1-\delta}\} \\ \forall x_{ij} \geq 0 \text{ and } x_{ij} \text{ is an integer} \end{cases}$$
(4)

In the above differential privacy constraints (each user log generates a constraint): due to $\forall t_{ijk} = \frac{c_{ij}}{c_{ij} - c_{ijk}} > 1$, the coefficient of all the linear constraints $\forall \log t_{ijk}$ should be greater than 0 (all unique query-url pairs have been removed). Letting $Mx \leq b$ be the above differential privacy constraints, all the elements in the constraint matrix $M$ are non-negative and all the elements in $b$ are equal to $\min\{\epsilon, \log \frac{1}{1-\delta}\}$. Thus, we have:

STATEMENT 1. *Differential privacy constraints (Equation 4) are always feasible and bounded.*

We show the above property from the geometric perspective of linear constraints. Specifically, linear constraints $\{Mx \leq b, x \geq 0, b > 0\}$ form a **convex polytope**, which is always feasible and bounded if $M, b \geq 0$ [26]. i.e. in Figure 2(a) (two differential privacy constraints are generated by two user logs which includes three distinct query-url pairs), the feasible region of $\{Mx \leq b, x \geq$

---
[1]The optimization problems result from any two neighboring inputs (especially the large neighboring inputs) generate similar optimal solutions. Thus, if $d$ is not too small, the output count difference can be bounded by $d$. Otherwise, if $d$ is required to be sufficiently small (for reducing sensitivity/noise), we remove some user logs (that cause large differences in two optimal solutions). This allows us to trade off utility for end-to-end differential privacy.

[2]For all unique query-url pairs, we let the output count be 0 (for satisfying Condition 1 in Theorem 1).

$0, b > 0\}$ is formed as polytope OABCDE by two constraints (the space below planes AFH and GCD). Similarly, in Figure 2(b) (three differential privacy constraints are generated by three user logs which includes two distinct query-url pairs), all the solutions in the feasible region OABC (the region below AD, FC and EH) satisfy all the differential privacy constraints. For more variables and constraints, more hyperplanes would form the polytope that is still feasible and bounded [26].

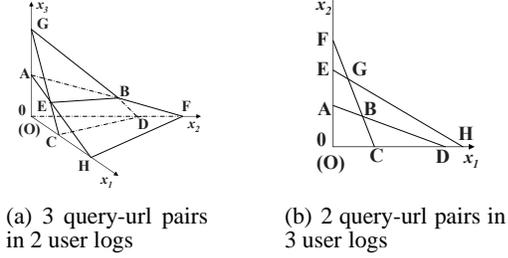

(a) 3 query-url pairs in 2 user logs  (b) 2 query-url pairs in 3 user logs

**Figure 2: Differential Privacy Constraints**

One interesting point worth noting is that the size of the output (the total number of all users' query-url pairs in the output) is bounded by the differential privacy constraints. If we regard the output size $\sum_{\forall (q_i,u_j)\in D} x_{ij}$ as the utility objective function, we can use the following problem to seek the optimal output utility:

$$\max : \sum_{\forall (q_i,u_j)\in D} x_{ij}$$

$$s.t. \begin{cases} \forall A_k \subset D, \sum_{\forall (q_i,u_j)\in A_k} x_{ij} \cdot \log t_{ijk} \leq \min\{\epsilon, \log \frac{1}{1-\delta}\} \\ \forall x_{ij} \geq 0 \text{ and } x_{ij} \text{ is an integer} \end{cases}$$

We define the above problem as *"Output size Utility-Maximizing Problem" (O-UMP)*. Since it is an integer linear programming (ILP) problem, we can solve it using some standard method (such as simplex algorithm) with linear relaxation [26] (the LP problem is always feasible and bounded). After solving it (optimal solution $x^* = \{\forall \lfloor x^*_{ij} \rfloor\}$), for every $(q_i, u_j)$, we sample user-IDs with $\lfloor x^*_{ij} \rfloor$ times multinomial trials (the input query-url-user histogram provides the probability of every sampled outcome in one trial). The sanitization algorithm satisfies Definition 2 (Proof in Appendix D).

LEMMA 1. *The O-UMP based sanitization algorithm satisfies $(\epsilon, \delta)$-probabilistic differential privacy for any pairs of neighboring input search logs.*

Since the optimal solution $x^* = \{\forall x^*_{ij}\}$ satisfies the differential privacy constraints, the randomization algorithm based on the linear relaxed solution should be also differentially private ($\forall \lfloor x^*_{ij} \rfloor \leq x^*_{ij}$, thus $\forall \lfloor x^*_{ij} \rfloor$ strictly satisfies the constraints $Mx \leq b$ where $M, b \geq 0$). Note that if we require adding Laplacian noise to $\{\forall x^*_{ij}\}$ to ensure differential privacy for the step of computing optimal counts, we cannot always guarantee that the noise-added optimal solution satisfies the differential privacy constraints, though this is likely (since the mean of added Laplacian noise is 0). Meanwhile, since the amount of noise $Lap(d/\epsilon')$ is directly proportional to $d$ (privacy parameter $\epsilon'$ is fixed), $d$ can be lowered to the preferred value (reducing the sensitivity/amount of noise) to gain closer approximation of strict end-to-end differential privacy. These also apply to the following utility-maximizing problems.

## 5.2 Optimal Utility of Frequent query-url Pairs

Top frequent click-through pairs in search logs have better utility [12] than abnormal query-url pairs for improving the quality of search results or enforcing the search with recommendations and suggestions. Retaining frequent query-url pairs in the sanitized search logs can be a basic and practical goal of seeking the optimal output utility in the sanitization. We denote this problem as *"Frequent query-url pair Utility-Maximizing Problem" (F-UMP)*.

First of all, we denote $|D|$ as the size (the total number of query-url pairs) of the input search log $D$. Thus, frequent query-url pairs can be identified using its *Support* in $D$: given a minimum support threshold $s$, if $\frac{c_{ij}}{|D|} \geq s$, then $(q_i, u_j)$ is a frequent click-through query-url pair in $D$. Since the support of a frequent query-url pair explicitly indicates its importance in the search log, the support of all the frequent query-url pairs should be preserved as much as possible. In other words, the support of every frequent query-url pair in the output $O$ should be close to its support in the input $D$ ($|D|$ does not include the number of unique query-url pairs which should be removed in the preprocessing step).

Thus, we can define the objective function as minimizing the sum of support distances for all the *"frequent query-url pairs"* in the input search log $D$:

$$\min : \sum_{\forall (q_i,u_j)\in D \text{ where } \frac{c_{ij}}{|D|}\geq s} ||\frac{x_{ij}}{|O|} - \frac{c_{ij}}{|D|}|| \quad (5)$$

where $|O| = \sum_{\forall (q_i,u_j)\in D} x_{ij}$ is the size of the output $O$.

With this objective, we formulate the F-UMP using the differential privacy constraints as below:

$$\min : \sum_{\forall (q_i,u_j)\in D \text{ where } \frac{c_{ij}}{|D|}\geq s} ||\frac{x_{ij}}{|O|} - \frac{c_{ij}}{|D|}||$$

$$s.t. \begin{cases} \forall A_k \subset D, \sum_{\forall (q_i,u_j)\in A_k} x_{ij} \cdot \log t_{ijk} \leq \min\{\epsilon, \log \frac{1}{1-\delta}\} \\ \sum_{\forall (q_i,u_j)\in D} x_{ij} = |O| \\ \forall x_{ij} \geq 0 \text{ and } x_{ij} \text{ is an integer} \end{cases}$$

Generally, since every query-url pair's support in $D$ and $O$ are two ratios, pursuing the minimized sum of support distances (our objective in F-UMP) cannot always guarantee an output with good frequent query-url pair utility (i.e. the number of all frequent query-url pairs are very small, but the support of them are close to the original one). Alternatively, we can specify a fixed output size $|O|$ in the sanitization and seek the optimal utility for the frequent query-url pairs. Recall that O-UMP can generate the output with the maximum size for any input $D$ and fixed parameters $(\epsilon, \delta)$ (*we denote the maximum output size as $\lambda$*). Thus, to preserve sufficient output size, we can solve the F-UMP with a specified constant output size $|O| \in (0, \lambda]$.

STATEMENT 2. *F-UMP can be considered as an integer linear programming (ILP) problem if we fix the output size $|O|$ as a constant and standardize the absolute values in the objective function.*

First, due to $|O| = \sum_{\forall (q_i,u_j)\in D} x_{ij}$, if we specify the size of the output in the sanitization, $\frac{x_{ij}}{|O|} - \frac{c_{ij}}{|D|}$ can be considered as linear. Second, we can transform the absolute values in the objective function in a standard way:

1. create a new variable $y_{ij}$ for every frequent query-url pair $\forall (q_i, u_j)$ where $\frac{c_{ij}}{|D|} \geq s$: $y_{ij} = \frac{x_{ij}}{|O|} - \frac{c_{ij}}{|D|}$;

2. generate two new constraints for every $y_{ij}$: $y_{ij} \geq \frac{x_{ij}}{|O|} - \frac{c_{ij}}{|D|}$ and $y_{ij} \geq \frac{c_{ij}}{|D|} - \frac{x_{ij}}{|O|}$.

As a result, F-UMP can be transformed into an integer linear programming (ILP) problem as below:

$$\min : \sum_{\forall (q_i,u_j) \in D \text{ where } \frac{c_{ij}}{|D|} \geq s} y_{ij}$$

$$s.t. \begin{cases} \forall A_k \subset D, \sum_{\forall (q_i,u_j) \in A_k} x_{ij} \cdot \log t_{ijk} \leq \min\{\epsilon, \log \frac{1}{1-\delta}\} \\ \sum_{\forall (q_i,u_j) \in D} x_{ij} = |O| \\ \forall (q_i,u_j) \text{ where } \frac{c_{ij}}{|D|} \geq s, y_{ij} \geq \frac{x_{ij}}{|O|} - \frac{c_{ij}}{|D|} \\ \forall (q_i,u_j) \text{ where } \frac{c_{ij}}{|D|} \geq s, y_{ij} \geq \frac{c_{ij}}{|D|} - \frac{x_{ij}}{|O|} \\ \forall x_{ij} \geq 0 \text{ and } x_{ij} \text{ is an integer} \end{cases}$$

Similar to O-UMP, we can solve the above ILP problem using some standard methods such as Simplex algorithm with linear relaxation [26] (if $|O|$ is specified to be no greater than $\lambda$, the ILP problem should be feasible and bounded).

Overall, in F-UMP based sanitization, we can specify an appropriate output size $|O| \in (0, \lambda]$, solve the ILP problem (optimal solution $x^* = \{\forall \lfloor x_{ij}^* \rfloor\}$) and generate the optimal output utility: the Input/Output Support of all the frequent query-url pairs tends to be close (only counting the non-unique query-url pairs) and the output size can be assured as well. Finally, we sample the output with the optimal solution of F-UMP: for every $(q_i, u_j)$ (either frequent or infrequent), we sample user-IDs with $\lfloor x_{ij}^* \rfloor$ times multinomial trials (equally, the input query-url-user histogram provides the probability of every sampled outcome in one trial). As discussed in Section 3.2, the shape of query-url-user histogram can be preserved in this problem based sanitization algorithm. Also, the sanitization algorithm satisfies Definition 2 (Proof in Appendix D).

LEMMA 2. *The F-UMP based sanitization algorithm satisfies $(\epsilon, \delta)$-probabilistic differential privacy for any pairs of neighboring input search logs.*

## 5.3 Maximizing query-url Pair Diversity

Occasionally, more distinct query-url pairs exhibit better utility, we can formulate the *"Diversity Utility-Maximizing Problem" (D-UMP)* in search log sanitization. The diversity of search logs normally has two facts: the diversity of search queries and the diversity of query-url pairs. Since we investigate the potential privacy breach from every query-url pair (finer-grained than search queries), we denote the diversity utility of search logs as the number of distinct query-url pairs. (Indeed, we can also model search query diversity maximizing problem in a similar way.)

In our sanitization, $x_{ij}$ represents the count of query-url pair $(q_i, u_j)$ in the output $O$. To evaluate the diversity of the sanitized search log $O$, we can introduce another variable $y_{ij}$ for every $x_{ij}$.

$$\begin{cases} y_{ij} = 1, & \text{if } x_{ij} > 0 \\ y_{ij} = 0, & \text{if } x_{ij} = 0 \end{cases} \quad (6)$$

We thus define the utility function as $\max : \sum y_{ij}$. Moreover, given a large constant $H \geq \max\{\forall c_{ij}\}$, Equation 6 is guaranteed to hold by the following inequalities:

$$\begin{cases} \forall (q_i, u_j), & x_{ij} \leq y_{ij} \cdot H \\ \forall (q_i, u_j), & x_{ij} \geq y_{ij} \\ y_{ij} \in \{0,1\}, & \forall x_{ij} \geq 0, H \geq max\{\forall c_{ij}\} \end{cases} \quad (7)$$

As a result, D-UMP can be formally defined as:

$$\max : \sum_{\forall (q_i,u_j) \in D} y_{ij}$$

$$s.t. \begin{cases} \forall A_k \subset D, \sum_{\forall (q_i,u_j) \in A_k} x_{ij} \cdot \log t_{ijk} \leq \min\{\epsilon, \log \frac{1}{1-\delta}\} \\ \forall (q_i,u_j) \in D, x_{ij} \leq y_{ij} \cdot H \\ \forall (q_i,u_j) \in D, x_{ij} \geq y_{ij} \\ H \geq max\{\forall c_{ij}\}, \forall x_{ij} \geq 0 \text{ and is an integer}, y_{ij} \in \{0,1\} \end{cases}$$

Essentially, letting $\forall x_{ij} \in \{0,1\}$ and $x_{ij} = y_{ij}$, the above mixed integer programming (MIP) problem can be transformed to a simplified binary integer programming (BIP) problem (see Equation 8). Both problems have the same optimal solution for variables $y = \{\forall y_{ij}\}$. (We prove Theorem 2 in Appendix C)

THEOREM 2. *The optimal solution $y^* = \{\forall y_{ij}^*\}$ of the BIP problem is equivalent to the values $\{\forall y_{ij}^*\}$ in the optimal solution $\{x^*, y^*\} = \{\forall x_{ij}^*, \forall y_{ij}^*\}$ of the MIP problem.*

$$\max : \sum_{\forall (q_i,u_j) \in D} y_{ij}$$

$$s.t. \begin{cases} \forall A_k \subset D, \sum_{\forall (q_i,u_j) \in A_k} y_{ij} \log t_{ijk} \leq \min\{\epsilon, \log \frac{1}{1-\delta}\} \\ H \geq max\{c_{ij}\}, \forall y_{ij} \in \{0,1\} \end{cases} \quad (8)$$

After solving the simpler BIP problem rather than the MIP problem (both problems are feasible), we thus let $\forall (q_i, u_j) \in D, x_{ij} = y_{ij} \in \{0, 1\}$ be the optimal solution of D-UMP (sampling *user-IDs* in only one trial for every query-url pair in the output. Similarly, the input query-url-user histogram provides the probability of every sampled outcome in one trial).

However, both BIP and MIP problem are NP-hard [26]. For large-scale D-UMP, we propose an effective and efficient heuristic algorithm to solve the BIP problem in Algorithm 2. It seeks an approximate optimal value for the BIP problem. We iteratively remove sensitive query-url pairs (let $y_{ij} = 0$ if $y_{ij}$ has a maximum positive coefficient $t_{ijk}$ in the sparse constraint matrix). We eliminate these query-url pairs since they belong to a certain user with the highest percent in the count histogram of the triplets query-url-user (sensitive to the corresponding user. i.e. if user $s_k$ holds 90% of $(q_i, u_j)$, $t_{ijk}$ should be large). The algorithm terminates until all the differential privacy constraints are satisfied.

---

**Algorithm 2** Sensitive query-url Pair Eliminating (SPE) Heuristic

**Input:** search log $D$ and differential privacy parameters $(\epsilon, \delta)$
**Output:** optimal solution for D-UMP $y^* = \{\forall y_{ij}^*\}$
1: remove all the unique query-url pairs from $D$ (preprocessing).
2: **for** every $(q_i, u_j) \in D$ **do**
3: $\quad y_{ij} \leftarrow 1$.
4: **while** true **do**
5: $\quad$ find the maximum $t_{ijk} = \frac{c_{ij}}{c_{ij} - c_{ijk}}$ from the constraint matrix.
6: $\quad$ let $y_{ij} \leftarrow 0$ for the maximum $t_{ijk}$.
7: $\quad$ **if** $\forall A_k, \sum_{\forall (q_i,u_j) \in A_k} y_{ij} \log t_{ijk} \leq \min\{\epsilon, \log \frac{1}{1-\delta}\}$ **then**
8: $\quad\quad$ break
9: return $y^* = \{\forall y_{ij}^*\}$.

---

The sanitization algorithm based on D-UMP also satisfies Definition 2. (Proof in Appendix D)

LEMMA 3. *The D-UMP based sanitization algorithm satisfies $(\epsilon, \delta)$-probabilistic differential privacy for any pairs of neighboring input search logs.*

## 6. EXPERIMENTAL RESULTS

### 6.1 Experiment Setup[3]

---

[3]Since the published search logs in [19] and [10] do not include pseudonymous user-IDs for associating distinct query-url pairs in every user's search history, the utility of our sanitized search logs is incomparable with their work. Moreover, since Laplacian noise has been well evaluated in their work, we focus on testing the optimal utility w.r.t. the output counts of all query-url pairs.

**Dataset.** In our experiments, we utilize the AOL real search log [3, 11] to test our utility-maximizing problems. Our experimental dateset is extracted from one subset of AOL data. Specifically, we randomly pick 2500 out of over 65000 user logs in the selected AOL data. We remove all the unique query-url pairs (appear in only one user log) from the selected dataset in our preprocessing step. Thus, Table 3 presents the characteristics of the AOL dataset (only collect the tuples with clicks), our randomly selected dataset and the preprocessed dataset. 6043 distinct query-url pairs is held by 1980 users in the preprocessed dataset (since search logs which are extremely diverse include large number of unique query-url pairs, most of the existing work [19, 10] cannot maintain the entire output diversity either). Thus, we have *6043 variables* and *1980 differential privacy constraints* in our UMPs.

**Table 3: Characteristics of the Data Sets**

|  | AOL Dataset | Exp. Dataset | Preprocessed Dataset (without unique pairs) |
|---|---|---|---|
| # of total tuples (size) | 1,864,860 | 237,786 | 53,067 ($|D|$) |
| # of user logs | 51,922 | 2,500 | **1,980** (Constraints) |
| # of distinct queries | 583,084 | 83,130 | 4,971 |
| # of distinct urls | 373,837 | 82,076 | 4,289 |
| # of query-url pairs | 1,190,491 | 163,681 | **6,043** (Variables) |

**Experimental Parameters Setup.** To observe the tuning of differential privacy parameters $(\epsilon, \delta)$, we let $\delta = \{10^{-4}, 10^{-3}, 10^{-2}, 10^{-1}, 0.2, 0.5, 0.8\}$ and $e^\epsilon = \{1.001, 1.01, 1.1, 1.4, 1.7, 2.0, 2.3\}$ in all three utility-maximizing problems. Furthermore, F-UMP requires two additional parameters: the minimum support $s$ and the output size $|O|$ ($|O| \leq \lambda$ and $\lambda$ is given as the optimal value of O-UMP). We let $s = \{\frac{1}{100}, \frac{1}{250}, \frac{1}{500}, \frac{1}{750}, \frac{1}{1000}\}$. For every pair of $\epsilon$ and $\delta$, we compute $\lambda$ in O-UMP and specify an appropriate output size $|O|$ in F-UMP.

**Experimental Platform.** All the experiments are performed on an HP machine with Intel Core 2 Duo CPU 3GHz and 3G RAM running Microsoft Windows XP Professional Operating system. While solving D-UMP, we also submit the AMPL format of the BIP problems to three NEOS solvers (qsopt_ex, scip and feaspump [16]) running online in addition to locally running our heuristic.

## 6.2 Maximum Output Size $\lambda$

With the preprocessed dataset ($|D|$ = 53067 as shown in Table 3), we can compute the maximum output size $\lambda$ using O-UMP for a given pair of differential privacy parameters $(e^\epsilon, \delta)$. Table 4 presents the maximum output size (the optimal value of O-UMP) for different pairs of $(e^\epsilon, \delta)$ where O-UMP is solved by Matlab function *linprog*. To generate the output $O$, we can sample userIDs for every query-url pair according to the optimal solution (6043 variables/query-url pairs). $|O|$ can be maximized while the entire process satisfies $(\epsilon, \delta)$-differential privacy. We can obtain 7.08%-26.2% of the original size with the given parameters. Due to the highly diversity and sparseness of search log data, this percent of output size is sufficient good for differential privacy guaranteed sanitization algorithms.

**Table 4: Maximum Output Size $\lambda$ on $e^\epsilon$ and $\delta$ ($|D|$ = 53067)**

| $e^\epsilon \backslash \delta$ | $10^{-4}$ | $10^{-3}$ | $10^{-2}$ | $10^{-1}$ | 0.2 | 0.5 | 0.8 |
|---|---|---|---|---|---|---|---|
| 1.001 | 3759 | 4007 | 4007 | 4007 | 4007 | 4007 | 4007 |
| 1.01 | 3759 | 4007 | 4879 | 4879 | 4879 | 4879 | 4879 |
| 1.1 | 3759 | 4007 | 4891 | 8382 | 8382 | 8382 | 8382 |
| 1.4 | 3759 | 4007 | 4891 | 8874 | 10445 | 11419 | 11419 |
| 1.7 | 3759 | 4007 | 4891 | 8874 | 10445 | 12438 | 12438 |
| 2.0 | 3759 | 4007 | 4891 | 8874 | 10445 | 13088 | 13088 |
| 2.3 | 3759 | 4007 | 4891 | 8874 | 10445 | 13088 | 13901 |

## 6.3 Optimal Utility of Frequent query-url Pairs

Recall that F-UMP based sanitization generates outputs with the minimum sum of the support distances of all the *frequent query-url pairs*. Thus, we examine the maximum frequent query-url pairs utility with three measures: the optimal value of F-UMP (minimum sum of the support distances, see Equation 5), the *Precision* and *Recall* of the frequent query-url pairs in the input/output. *Precision* and *Recall* are defined as below:

$$Precision = \frac{|S_0 \cap S|}{|S|}, Recall = \frac{|S_0 \cap S|}{|S_0|} \quad (9)$$

where $S_0$ and $S$ denote the set of frequent query-url pairs in $D$ and $O$ respectively, and $|\cdot|$ means the cardinality of the set. Specifically, *Precision* is defined to evaluate the fraction of the frequent query-url pairs in the output that are originally frequent in the input with the same minimum support. *Recall* is defined to evaluate the fraction of the frequent query-url pairs in the input that remains frequent in the output with the same minimum support.

To evaluate the performance of F-UMP in differentially private search log sanitization, we run two groups of experiments. First, we fix the output size and the minimum support as: $|O| = 3000 < \lambda$ and $s = \frac{1}{500}$, and test the (measurement) results with different pairs of $(\epsilon, \delta)$. Second, we fix the differential privacy parameters as: $e^\epsilon = 2, \delta = 0.5$ ($\lambda = 13088$, as shown in Table 4), and test the results with different minimum support $s$ and output size $|O|$. One essential point worth noting is that the minimum sum of support distances is an effective measure in the first group of experiments because the minimum support $s$ is fixed and the original frequent query-url pairs in the input has been determined for all different pairs of $\epsilon$ and $\delta$ (thus the sum of the support distances for all the frequent query-url pairs in the input is comparable). However, in the second group, the set of original frequent query-url pairs is varying for different $s$, hence the objective values of F-UMP is incomparable on a varying $s$. Therefore, we use the average of the support distances for all the frequent query-url pairs in the input in addition to the sum of them in the second group of experiments.

Interestingly, in all our F-UMP experiments, *Precision* is always equal to 1, which means all the frequent query-url pairs in the output are also frequent in the input with the same minimum support $s$. This is quite reasonable: suppose that $(q_i, u_j)$ is not a frequent query-url pair in the input where $\frac{c_{ij}}{|D|} < s$, if it is frequent in the output where $\frac{x_{ij}}{|O|} \geq s$, the solution of F-UMP must be not optimal (reducing $\frac{x_{ij}}{|O|}$ to $\frac{c_{ij}}{|D|}$ might improve the objective value and does not violate differential privacy constraints).

In the first group of experiments, Figure 3(a) and 3(b) demonstrate the *Recall* and *Sum of the Support Distances* for all the frequent query-url pairs in the input. Fixing $\delta$, *Recall* increases as $\epsilon$ increases until $\epsilon = \log \frac{1}{1-\delta}$. Fixing $\epsilon \geq \log \frac{1}{1-\delta}$, *Recall* increases as $\delta$ increases; fixing $\epsilon < \log \frac{1}{1-\delta}$, *Recall* stays invariant even if $\delta$ is increasing. By contrast, the sum of support distances has an inverse increasing trend on varying $\epsilon$ and $\delta$.

**Table 5:** *Recall* **on Output Size $|O|$ and Minimum Support $s$** ($e^\epsilon = 2, \delta = 0.5, \lambda = 13088$)

| $s \backslash |O|$ | 3000 | 4000 | 5000 | 6000 | 7000 | 8000 |
|---|---|---|---|---|---|---|
| $\frac{1}{1000}$ | 0.8873 | 0.8189 | 0.874 | 0.8661 | 0.8583 | 0.8346 |
| $\frac{1}{750}$ | 0.8095 | 0.8762 | 0.8571 | 0.8476 | 0.8952 | 0.8667 |
| $\frac{1}{500}$ | 0.9143 | 0.9143 | 0.9286 | 0.9143 | 0.8857 | 0.8714 |
| $\frac{1}{250}$ | 0.9116 | 0.8529 | 0.8529 | 0.8529 | 0.8529 | 0.8235 |
| $\frac{1}{100}$ | 0.933 | 0.8667 | 0.8 | 0.8 | 0.8 | 0.7333 |

In the second group of experiments, Table 5 presents the *Recall* on different pairs of outputs size and minimum support. As we

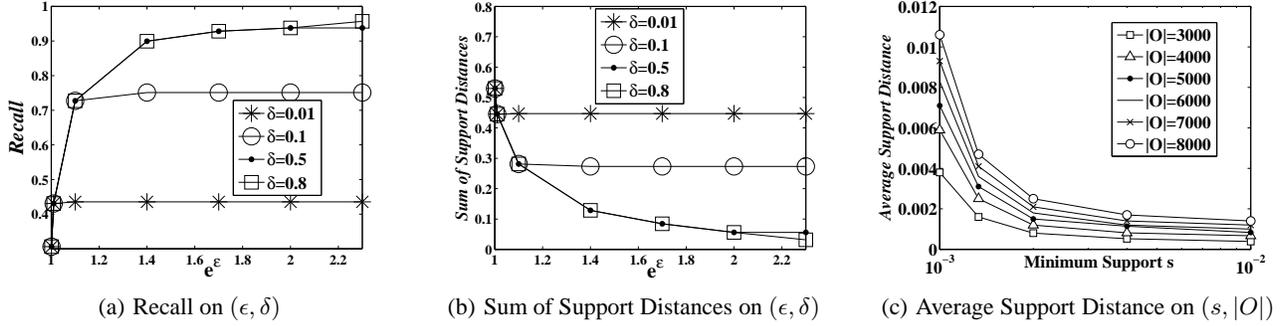

(a) Recall on $(\epsilon, \delta)$  (b) Sum of Support Distances on $(\epsilon, \delta)$  (c) Average Support Distance on $(s, |O|)$

**Figure 3: F-UMP Performance**

can see, over 80% of the frequent query-url pairs can be retained in the output with fixing $e^\epsilon = 2$ and $\delta = 0.5$ (given more strict $e^\epsilon$ and $\delta$, 30% of them can be retained as shown in Figure 3(a)). In addition, Table 6 illustrates the sum of support distances *for all frequent query-url pairs in the input* (the same $|O|$ and $s$ as Table 5). Fixing $s$, the sum of support distances increases as the output size increases (they are comparable due to fixed $s$). This fact is true: given a fixed minimum support $s$, for the fixed set of frequent query-url pairs in the input, it is easier to achieve the minimum support without violating differential privacy constraints when $|O|$ is not too large (the ideal output count $x_{ij}$ is $|O| \cdot \frac{c_{ij}}{|D|}$ and the output counts are bounded by privacy constraints, thus all frequent query-url pairs $\forall x_{ij}$ are likely to achieve $|O| \cdot \frac{c_{ij}}{|D|}$ if $|O|$ is small). Finally, since the set of frequent query-url pairs varies for different $s$, we compare the average support distance instead of the sum of them for different $s$. As shown in Figure 3(c), the average support distance decreases as the minimum support $s$ increases (logarithmic scale minimum support $s$). Therefore, the frequent query-url pairs in the output is closer to them in the input if a larger minimum support is given in the F-UMP.

**Table 6: *Sum of Freq. query-url Pair Support Distances* on Output Size $|O|$ and Min. Support $s$ ($e^\epsilon = 2, \delta = 0.5, \lambda = 13088$)**

| $s \backslash |O|$ | 3000 | 4000 | 5000 | 6000 | 7000 | 8000 |
|---|---|---|---|---|---|---|
| $\frac{1}{1000}$ | 0.0551 | 0.085 | 0.1058 | 0.1279 | 0.1485 | 0.1785 |
| $\frac{1}{750}$ | 0.0549 | 0.0854 | 0.1116 | 0.1271 | 0.1477 | 0.1767 |
| $\frac{1}{500}$ | 0.0559 | 0.0865 | 0.1048 | 0.1247 | 0.1448 | 0.1716 |
| $\frac{1}{250}$ | 0.0555 | 0.086 | 0.1043 | 0.1236 | 0.1393 | 0.161 |
| $\frac{1}{100}$ | 0.0574 | 0.088 | 0.1063 | 0.1246 | 0.1392 | 0.1583 |

## 6.4 Maximum query-url Pair Diversity

### 6.4.1 D-UMP Performance

We now look at the performance of D-UMP (maximum diversity utility). Figure 4 shows the percentage of retained query-url pairs in the output with the same parameters $(\epsilon, \delta)$ as F-UMP. The maximum query-url diversity has a similar increasing trend as the *Recall* of F-UMP (Figure 3(a)). Moreover, the query-url diversity can be retained as high as 30%. Note: the input has been preprocessed by removing all the unique query-url pairs, and they are not counted in the denominator of the ratio.

### 6.4.2 BIP Solver Comparison

Since D-UMP is an NP-hard problem, we introduced an effective heuristic algorithm (Algorithm 2) for this binary integer programming (BIP) problem with a sparse non-negative constraint matrix. We now compare the performance of our Sensitive Pair Eliminating heuristic (SPE) with some popular BIP solvers (Matlab bintprog function, Neos qsopt_ex, Neos scip and Neos feaspump [16]).

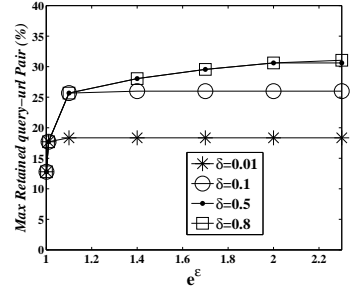

**Figure 4: Maximum Diversity on $(\epsilon, \delta)$ (Algorithm 2)**

**Table 7: Retained Diversity Utility of Different BIP Solvers**

(a) $e^\epsilon = 2$

| BIP Solver $\backslash \delta$ | $10^{-3}$ | $10^{-2}$ | $10^{-1}$ | 0.2 | 0.5 | 0.8 |
|---|---|---|---|---|---|---|
| SPE (Heuristic) | **12.8%** | **18.1%** | **26.0%** | **28.1%** | 29.5% | **30.6%** |
| Matlab bintprog | 9.6% | 15.2% | 23.8% | 26.8% | 28.9% | 29.5% |
| Neos qsopt_ex | 9.6% | 15.2% | 23.4% | 26.8% | 29.5% | 29.5% |
| Neos scip | 9.5% | 15.2% | 23.7% | 26.8% | 29.5% | 29.5% |
| Neos feaspump | 9.6% | 15.2% | 25.8% | 29.5% | **30.3%** | 30.3% |

(b) $\delta = 0.1$

| BIP Solver $\backslash e^\epsilon$ | 1.01 | 1.1 | 1.4 | 1.7 | 2.0 | 2.3 |
|---|---|---|---|---|---|---|
| SPE (Heuristic) | **17.7%** | 25.7% | **26.0%** | **26.0%** | **26.0%** | **26.0%** |
| Matlab bintprog | 14.6% | 22.5% | 23.8% | 23.8% | 23.8% | 23.8% |
| Neos qsopt_ex | 15.5% | 22.5% | 23.4% | 23.4% | 23.4% | 23.4% |
| Neos scip | 14.6% | 21.4% | 23.1% | 23.1% | 23.1% | 23.1% |
| Neos feaspump | 15.5% | **25.8%** | 25.8% | 25.8% | 25.8% | 25.8% |

As shown in Table 7, we collected the maximum percent of retained distinct query-url pairs using all the solvers with the same experimental inputs. We observe that our heuristic algorithm performs better than other solvers in most cases and the optimal values by all the solvers have quite similar varying tendency. Specifically, Algorithm 2 generates sanitized search logs with greater query-url pair diversity than Matlab bintprog, NEOS qsopt and Neos scip. NEOS feaspump performs slightly better than Algorithm 2 only when $(e^\epsilon = 2, \delta = 0.5)$ and $(e^\epsilon = 1.1, \delta = 0.1)$.

Finally, we plot the computational costs for solving a typical D-UMP by all solvers in Figure 5 ($e^\epsilon = 1.7, \delta = 10^{-3}$). Since our Sensitive query-url Pair Eliminating (SPE) algorithm has the complexity $O(n^2 \log mn)$ (constraint matrix size: $m \times n$), it outperforms other solvers for our D-UMP in time complexity as well.

## 6.5 Difference of Input/Output Histograms

As described in Section 3.2, our multinomial sampling, particularly the F-UMP based sanitization can retain the shape of the histograms in the output (generate similar count histograms for distinct triplets: query-url-user $(q_i, u_j, s_k)$). We now examine this by comparing two histograms.

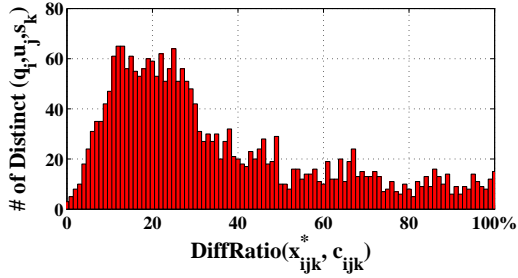
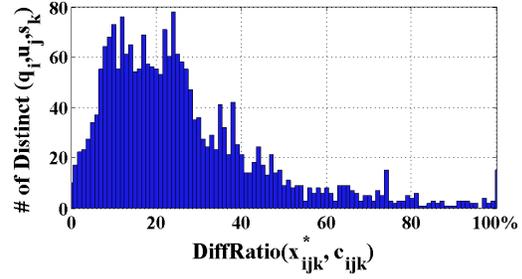

(a) |O|=4000

(b) |O|=6000

**Figure 6: The Difference Ratio of Input and Output query-url-user (Triplets) Histogram (F-UMP based Sanitization)**

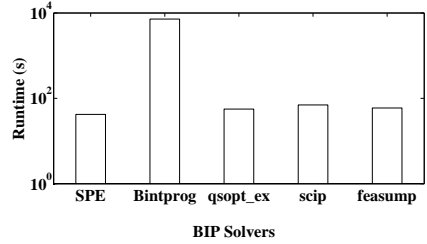

**Figure 5: Computational Performance for Solving D-UMP ($e^\epsilon = 1.7, \delta = 10^{-3}$, Logarithmic scale runtime)**

Specifically, we generate 10 randomized outputs according to the optimal solution of F-UMP for two different output size $|O| = 4000$ and 6000 respectively (fixing $e^\epsilon = 2, \delta = 0.5, s = 1/500$), and plot two bar plots in Figure 6: the X-axis varies from 0% to 100% while the Y-axis represents *the average number of distinct triplets $(q_i, u_j, s_k)$[4] whose difference ratio of the input/output histograms (defined in Equation 10) equals the values in the X-axis*. In both Figure 6(a) and 6(b), the percent of most triplets $(q_i, u_j, s_k)$ in the input/output varies within a tolerable bound ($|O| = 4000$, the difference ratio of about 75% triplets is below 40%; $|O| = 6000$, the difference ratio of about 90% triplets is below 40%).

$$DiffRatio(x^*_{ijk}, c_{ijk}) = ||\frac{x^*_{ijk}/|O| - c_{ijk}/|D|}{c_{ijk}/|D|}|| \quad (10)$$

## 7. CONCLUSION AND FUTURE WORK

In this paper, we have addressed the important practical problem of retaining the maximum utility while the search log sanitization satisfies differential privacy and generates outputs with the identical schema as the original search log. As a necessary step, we have defined three different notions of utility that are useful for various applications. We have implemented our approach and validated it on several real data sets.

We can extend our work in several directions. First, additional notions of utility can be considered and corresponding optimization models created. We also need to explore ways of combining different utility notions to create a single joint objective. This would be akin to a multi-objective optimization. Second, corresponding to the utility-maximizing problem, one can similarly define the privacy breach-minimizing problem which asks for minimal privacy loss while satisfying a certain utility. Third, since we have modeled the utility-maximizing problems in the optimization framework, it should be possible to leverage the significant work in the field of operations research to solve these problems. We intend to explore these in the future.

## 8. REFERENCES

[1] E. Adar. User 4xxxxx9: Anonymizing query logs. In *Workshop at the WWW '07*, Banff, Alberta, Canada, May 2007. WWW.
[2] K. Baclawski. *Introduction to probability with R*. Chapman and Hall/CRC, 1 edition, 2008.
[3] M. Barbaro and T. Zeller. A face is exposed for aol searcher no. 4417749, August 9, 2006. (New York Times).
[4] R. Bayardo and R. Agrawal. Data privacy through optimal k-anonymization. In *ICDE*, pages 217–228, 2005.
[5] A. Cooper. A survey of query log privacy-enhancing techniques from a policy perspective. *TWEB*, 2(4), 2008.
[6] C. Dwork, K. Kenthapadi, F. McSherry, I. Mironov, and M. Naor. Our data, ourselves: Privacy via distributed noise generation. In *EUROCRYPT*, pages 486–503, 2006.
[7] C. Dwork, F. McSherry, K. Nissim, and A. Smith. Calibrating noise to sensitivity in private data analysis. In *TCC*, pages 265–284, 2006.
[8] A. Ghosh, T. Roughgarden, and M. Sundararajan. Universally utility-maximizing privacy mechanisms. In *STOC '09*, pages 351–360, New York, NY, USA, 2009. ACM.
[9] J. Ginsberg, M. H. Mohebbi, R. S. Patel, L. Brammer, M. S. Smolinski, and L. Brilliant. Detecting influenza epidemics using search engine query data. *Nature*, 457:1012 – 1014, Feb. 2009.
[10] M. Götz, A. Machanavajjhala, G. Wang, X. Xiao, and J. Gehrke. Publishing search logs - a comparative study of privacy guarantees. *IEEE TKDE, to appear*.
[11] K. Hafner. Researchers yearn to use aol logs, but they hesitate, August 23, 2006. (New York Times).
[12] J. Han, H. Cheng, D. Xin, and X. Yan. Frequent pattern mining: current status and future directions. *DMKD*, 15(1):55–86, 2007.
[13] M. Hay, C. Li, G. Miklau, and D. Jensen. Accurate estimation of the degree distribution of private networks. In *ICDM 2009*, pages 169–178.
[14] Y. He and J. F. Naughton. Anonymization of set-valued data via top-down, local generalization. *PVLDB*, 2(1):934–945, 2009.
[15] Y. Hong, X. He, J. Vaidya, N. R. Adam, and V. Atluri. Effective anonymization of query logs. In *CIKM*, pages 1465–1468, 2009.
[16] M. M. J. Czyzyk and J. Morĺe. The neos server.
[17] R. Jones, R. Kumar, B. Pang, and A. Tomkins. "i know what you did last summer": query logs and user privacy. In *CIKM '07*, pages 909–914, 2007.
[18] D. Kifer and J. Gehrke. Injecting utility into anonymized datasets. In *SIGMOD Conference*, pages 217–228, 2006.
[19] A. Korolova, K. Kenthapadi, N. Mishra, and A. Ntoulas. Releasing search queries and clicks privately. In *WWW*, pages 171–180, 2009.

---

[4]The triplets w.r.t. infrequent query-url pairs can be ignored in general. If $s$ is sufficiently small, the shape of the query-url-user histogram w.r.t. *all query-url pairs* can be optimally retained.

[20] R. Kumar, J. Novak, B. Pang, and A. Tomkins. On anonymizing query logs via token-based hashing. In *WWW*, pages 629–638, 2007.
[21] K. LeFevre, D. J. DeWitt, and R. Ramakrishnan. Mondrian multidimensional k-anonymity. *ICDE 2006*, page 25.
[22] T. Li and N. Li. On the tradeoff between privacy and utility in data publishing. In *KDD*, pages 517–526, 2009.
[23] J. Liu and K. Wang. Enforcing vocabulary k-anonymity by semantic similarity based clustering. In *ICDM*, pages 899–904, 2010.
[24] A. Machanavajjhala, D. Kifer, J. M. Abowd, J. Gehrke, and L. Vilhuber. Privacy: Theory meets practice on the map. In *ICDE*, pages 277–286, 2008.
[25] F. McSherry and I. Mironov. Differentially private recommender systems: building privacy into the net. In *KDD 2009*, pages 627–636.
[26] C. H. Papadimitriou and K. Steiglitz. *Combinatorial optimization: algorithms and complexity*. Prentice-Hall, Inc., NJ, USA, 1982.
[27] X. Xiao, G. Wang, and J. Gehrke. Differential privacy via wavelet transforms. In *ICDE*, pages 225–236, 2010.

# APPENDIX

## A. PROOF OF THEOREM 1

PROOF. Assume that $D$ and $D'$ differ in an arbitrary user $s_k$'s user log $A_k$. In Section 4.1, we discussed two sets of output spaces $\Omega = \Omega_1 \cup \Omega_2$: all the possible outputs in $\Omega_1$ include $s_k$ whereas all the possible outputs in $\Omega_2$ does not include $s_k$. Hence, if the probabilities inequalities in Definition 2 hold for the above $\Omega_1, \Omega_2$, $(\epsilon, \delta)$-probabilistic differential privacy can be guaranteed for the randomization algorithm with this output space split.

First, according to Equation 2, if $\forall A_k \subset D, 1 - \prod_{\forall(q_i,u_j)\in A_k} (\frac{c_{ij}-c_{ijk}}{c_{ij}})^{x_{ij}} \leq \delta$ (Condition 3) holds, we have $Pr[\mathcal{R}(D) \in \Omega_1] \leq \delta$ for any input $D$. Meanwhile, Condition 1 guarantees that $Pr[\mathcal{R}(D) \in \Omega_1]$ can be effectively bounded by $\delta$. Otherwise, if a unique query-url pair$(q_i, u_j)$, given $x_{ij} > 0$, $Pr[\mathcal{R}(D) \in \Omega_1]$ should be equal to 1 with such output space split (no other space split available for any pair of neighboring input search logs).

Second, for all $O \in \Omega_2$, we have $Pr[\mathcal{R}(D') = O] > 0$ and $Pr[\mathcal{R}(D) = O] > 0$. If $D' \subset D$, Condition 2 ensures $\frac{Pr[\mathcal{R}(D)=O]}{Pr[\mathcal{R}(D')=O]} \leq 1 \leq \frac{Pr[\mathcal{R}(D')=O]}{Pr[\mathcal{R}(D)=O]} \leq e^\epsilon$. On the contrary, if $D \subset D'$, Condition 2 derived from $D'$ can also guarantees $\frac{Pr[\mathcal{R}(D')=O]}{Pr[\mathcal{R}(D)=O]} \leq 1 \leq \frac{Pr[\mathcal{R}(D)=O]}{Pr[\mathcal{R}(D')=O]} \leq e^\epsilon$.

Thus, the randomization algorithm $\mathcal{R}$ satisfies $(\epsilon, \delta)$-probabilistic differential privacy (by dividing output space as above) if three conditions in the theorem hold. Note that the violation of any condition would result in unbounded multiplicative and/or additive probability difference (given $\epsilon$ and $\delta$) for at least one input $D$ and/or one of its neighboring input $D'$ (Differential privacy will not be guaranteed), then the upper bounds $\epsilon$ and $\delta$ are tight. □

## B. PROOF OF PROPOSITION 1

PROOF. W.o.l.g., assume that two arbitrary neighboring search logs $D$ and $D'$ differing in one user log: $D = D' + A_k$ and $\widehat{O} \subseteq \Omega$ is an arbitrary set of possible outputs. For any input $D$, we can divide the output space $\Omega$ into two sets $\Omega_1$ and $\Omega_2$, such that (1) $Pr[\mathcal{R}(D) \in \Omega_1] \leq \delta$, and for $D, D'$ (2) $\forall O \in \Omega_2, 1/e^\epsilon \leq \frac{Pr[\mathcal{R}(D')=O]}{Pr[\mathcal{R}(D)=O]} \leq e^\epsilon$.

Let $\widehat{O}_1 = \widehat{O} \cap \Omega_1$ and $\widehat{O}_2 = \widehat{O} \cap \Omega_2$, thus: $Pr[\mathcal{R}(D) \in \widehat{O}] = \int_{\forall O \in \widehat{O}_1} Pr[\mathcal{R}(D) = O]dO + \int_{\forall O \in \widehat{O}_2} Pr[\mathcal{R}(D) = O]dO$
$\leq \int_{\forall O \in \Omega_1} Pr[\mathcal{R}(D) = O]dO + e^\epsilon \int_{\forall O \in \widehat{O}_2} Pr[\mathcal{R}(D') = O]dO$
$\leq \delta + e^\epsilon \int_{\forall O \in \widehat{O}_2} Pr[\mathcal{R}(D') = O]dO$
$\leq \delta + e^\epsilon Pr[\mathcal{R}(D') \in \widehat{O}_2] \leq \delta + e^\epsilon Pr[\mathcal{R}(D') \in \widehat{O}]$.

Similarly, we can prove that $Pr[\mathcal{R}(D') \in \widehat{O}] \leq \delta + e^\epsilon Pr[\mathcal{R}(D) \in \widehat{O}]$.

This completes the proof. □

## C. PROOF OF THEOREM 2

PROOF. To distinguish two optimal solutions $y^*$ in the BIP and the MIP problem, we denote $y^*$ for the BIP and the MIP problem as $(y^*)_B = \{\forall(y^*_{ij})_B\}$ and $(y^*)_M = \{\forall(y^*_{ij})_M\}$.

- Suppose that $\exists(y^*_{ij})_B = 0, (y^*_{ij})_M = 1$ and $\forall z \neq ij, (y^*_z)_B = (y^*_z)_M$ ($(y^*)_B$ and $(y^*)_M$ differ in one variable). Due to $(y^*_{ij})_M = 1$ and $x^*_{ij} \geq (y^*_{ij})_M$, all the constraints $\forall A_k \subset D, \sum_{\forall(q_i,u_j)\in A_k}(y_{ij})_M \cdot \log t_{ijk} \leq \min\{\epsilon, \log \frac{1}{1-\delta}\}$ must be satisfied for $(y^*)_M$.
  In addition, $(y^*_{ij})_M > (y^*_{ij})_B \implies \sum_{\forall(q_i,u_j)\in D}(y^*_{ij})_M > \sum_{\forall(q_i,u_j)\in D}(y^*_{ij})_B$. As $\forall(y^*_{ij})_B$ satisfies the constraints $\forall A_k \subset D, \sum_{\forall(q_i,u_j)\in A_k}(y_{ij})_B \cdot \log t_{ijk} \leq \min\{\epsilon, \log \frac{1}{1-\delta}\}$ in the BIP problem, $\sum_{\forall(q_i,u_j)\in D}(y^*_{ij})_M$ should be the optimal value for the BIP problem if other constraints are the same for two problems (due to $\sum_{\forall(q_i,u_j)\in D}(y^*_{ij})_M > \sum_{\forall(q_i,u_j)\in D}(y^*_{ij})_B$). Hence, it is a contradiction.

- Suppose that $\exists(y^*_{ij})_B = 1, (y^*_{ij})_M = 0$ and $\forall z \neq ij, (y^*_z)_B = (y^*_z)_M$ ($(y^*)_B$ and $(y^*)_M$ differ at one variable). Hence, the constraints $\forall A_k \subset D, \sum_{\forall(q_i,u_j)\in A_k}(y_{ij})_B \log t_{ijk} \leq \min\{\epsilon, \log \frac{1}{1-\delta}\}$ are satisfied in the BIP problem. In the MIP problem, if letting $x_{ij}$ be 1 for all $(y^*_{ij})_B = 1, \forall A_k \subset D, \sum_{\forall(q_i,u_j)\in A_k} x_{ij} \log t_{ijk} \leq \min\{\epsilon, \log \frac{1}{1-\delta}\}$ can be equally satisfied. In this case, we have $\sum_{\forall(q_i,u_j)\in D}(y^*_{ij})_B$
  $= \sum_{\forall(q_i,u_j)\in D} x_{ij} = \sum_{\forall(q_i,u_j)\in D}(y_{ij})_M$
  $> \sum_{\forall(q_i,u_j)\in D}(y^*_{ij})_M$
  (since $\forall(q_i, u_j) \in D, x_{ij} = (y_{ij})_M$). Hence, $(y^*)_M$ is not the optimal solution of the MIP problem. It is a contradiction.

Therefore, Theorem 2 has been proven. □

## D. PROOF OF LEMMA 1, 2 AND 3

PROOF. It is similar and straightforward to prove Lemma 1, 2 and 3 (probabilistic differential privacy) using Theorem 1, we thus prove them together.

The sanitized search log $O$ is generated in terms of the optimal solution of O-UMP, F-UMP or D-UMP. We sample the output based on the linear relaxed optimal solution $x^* = \{\lfloor x^*_{ij} \rfloor\}$ (gives the total count) and the query-url-user histograms in any input $D$ (gives the individual outcome probabilities). Due to $\forall \lfloor x^*_{ij} \rfloor \leq x^*_{ij}$, we can infer that $\{\forall(q_i, u_j) \in O, \lfloor x^*_{ij} \rfloor\}$ satisfies the Condition 2 and 3 of Theorem 1 (differential privacy constraints $\forall A_k \subset D, \sum_{\forall(q_i,u_j)\in A_k} x_{ij} \log t_{ijk} \leq \min\{\epsilon, \log \frac{1}{1-\delta}\}$ in O-UMP, F-UMP or D-UMP are satisfied). Moreover, Condition 1 of Theorem 1 is also guaranteed in the preprocessing step.

Thus, while sampling user-IDs for any input search log $D$ and its arbitrary neighboring input $D'$ with the optimal counts (given by the optimization problem), we can divide the multinomial sampling output space $\Omega$ (derived from $D$ and $D'$) into $\Omega_1$ and $\Omega_2$ as described in Section 4 where all the probabilities in Definition 2 are bounded by $\epsilon$ and $\delta$ in such space split (refer Theorem 1). Therefore, the O/F/D-UMP based sanitization (randomization) algorithm satisfies $(\epsilon, \delta)$-probabilistic differential privacy (we can add Laplacian noise to ensure differential privacy for the step of computing the optimal counts if necessary).

This completes the proof. □